# Area-efficient Selective Multi-Threshold CMOS Design Methodology for Standby Leakage Power Reduction

Takeshi Kitahara, Naoyuki Kawabe, Fimihiro Minami,
Katsuhiro Seta, and Toshiyuki Furusawa[*]

TOSHIBA Corporation Semiconductor Company    [*]TOSHIBA Microelectronics Corporation
580-1, Horikawa-cho, Saiwai-ku, Kawasaki, 212-8520, JAPAN
kitahara@dad.eec.toshiba.co.jp

**Abstract**

*This paper presents a design flow for an improved selective multi-threshold(Selective-MT) circuit. The Selective-MT circuit is improved so that plural MT-cells can share one switch transistor. We propose the design methodology from RTL(Register Transfer Level) to final layout with optimizing switch transistor structure.*

## 1. Introduction

The market for portable electric appliances has grown rapidly, generating great interest in low-power design. Not only dynamic power dissipation, but also standby leakage power has become a critical issue. The Dual-Vth(threshold voltage) technique[1] is an effective method to reduce the leakage power; however, low-Vth cells on critical paths still dissipate large leakage power, and this is limitation of the Dual-Vth technique.

For this reason, we developed the Selective-MT technique[2,3]. The technique uses the MT-cell[2,3], instead of the low-Vth cell. The MT-cell is less leaky than the low-Vth cell on standby. Selective-MT technique can save the leakage power more than the Dual-Vth technique can. One drawback of the conventional Selective-MT technique is the area overhead. The technique uses the MT-cell, and they are larger than the low-Vth cell. We have improved the technique to overcome the drawback, so that plural MT-cells can share one switch transistor. The critical path delay changes according to the switch cell structure, so novel P&R methodology is required to design the improved Selective-MT circuit.

## 2. Selective-MT Technique

The conventional Selective-MT technique uses the MT-cell. Basic structure of the MT-cell is shown in Fig.1(a). The logic part of the MT-cell consists of low-Vth transistors, and a high-Vth switch transistor controlled by the signal MTE is inserted. This switch transistor makes the MT-cell faster than the high-Vth cell and less leaky than the low-Vth cell. Fig.2 shows an example of the conventional Selective-MT circuit. MT-cells are used on critical paths, and high-Vth cells used on non-critical paths.

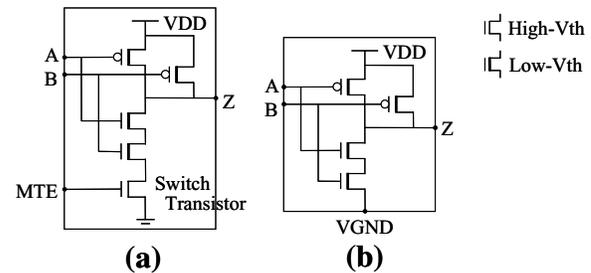

Fig.1: Basic Structure of 2-input NAND MT-cell

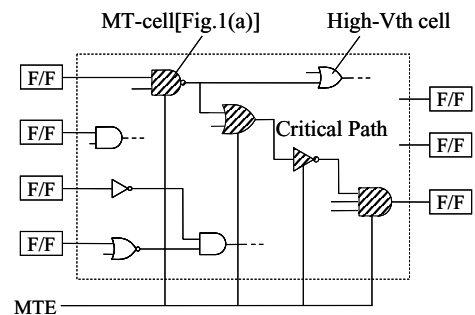

Fig.2: Conventional Selective-MT circuit

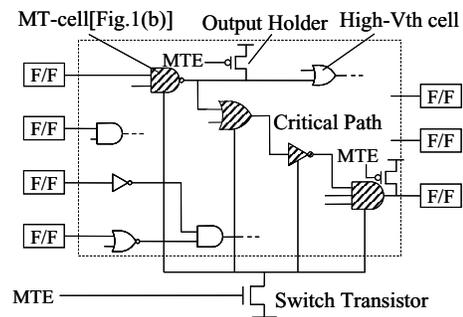

Fig.3: Improved Selective-MT circuit

To reduce the area overhead by MT-cells, we have separated the switch transistor from the MT-cell. Basic structure of an improved MT-cell is shown in Fig.1(b). The VGND(virtual ground) port is added to the MT-cell instead of the switch transistor, and the VGND port of the MT-cell is connected to the drain of the switch transistor. Not only the switch transistor, but also the output-hold circuit(output holder)[2] has been separated. The output holder sets the output of the improved MT-cell to one when a circuit is on standby, and it prevents a circuit from dissipating unex-pected power. Fig.3 shows an example of the improved Selective-MT circuit. The circuits in Fig.2 and Fig.3 are equivalent.

## 3. Design Methodology for Selective-MT

Fig.4 shows the design flow to generate the improved Selective-MT circuit. First, the RTL description is synthesized to generate an initial netlist and initial placement. Only low-Vth cells are used at this stage. As the low-Vth cell is faster, the timing constraint can be satisfied. Then, the circuit with low-Vth cells is replaced by high-Vth cells and MT-cells(without VGND ports) with the timing specification satisfied. The MT-cell(without VGND port) has the same information as the one(with VGND port), except that there is no definition related to the VGND port. The switch transistor does not appear at this stage, so the information related to the VGND port is unnecessary. MT-cells are assigned to gates on critical paths, and high-Vth cells on non-critical paths, as shown in Fig.3. This replacement is executed by the method which is similar to the way of generating the Dual-Vth circuit.

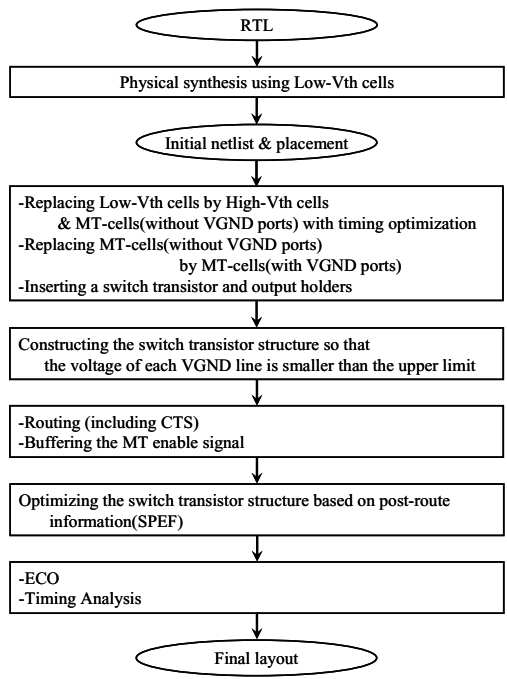

**Fig.4: Selective-MT design flow**

After a circuit with high-Vth cells and MT-cells(without VGND ports) is generated, a switch transistor and output holders are inserted. All MT-cells(without VGND ports) are replaced by the ones(with VGND ports) at this stage. One switch transistor is added, and all VGND ports at the MT-cells are connected to the drain of the switch transistor for generating an initial switch transistor structure. The output holder is inserted to the nets where it is required to prevent the signal floating. The output holder is not necessary for all MT-cells, as shown in Fig.3. When all fanouts of the MT-cell are connected to MT-cells, an output holder is unnecessary.

Next, the switch transistor structure is constructed. This task is executed by the back-end optimizer $CoolPower^{TM}$[4]. The tool generates clusters of MT-cells, and all VGND ports of MT-cells in one cluster are connected to the same switch transistor. It also decides the size of each switch transistor, so that the voltage bounce of each VGND line may not exceed the upper limit which the designer specifies. The switch transistor structure is constructed so that the wire length of each VGND line may not exceed an upper limit, as a long VGND line tends to suffer from the crosstalk. The number of MT-cell which shares the same switch transistor is also cared to prevent the electro-migration.

Then, routing including CTS(Clock Tree Synthesis) is executed. The MT enable signal MTE shown in Fig.3 has many fanouts, as MTE is necessary to be connected to all switch transistors and output holders. So, buffers needs to be inserted to the MTE net appropriately. When constructing the switch transistor structure before routing is done, the information about the resistance and the capacitance of each wire is estimated based on the placement information. So, there is an error when compared with the precise RC information which is generated after routing is done. After it is extracted, the re-optimization of the switch transistor structure is executed by $CoolPower^{TM}$ again. The size of each switch transistor is adjusted, so as that the voltage bounce of each VGND line may not exceed the upper limit. Finally, ECO(Engineering Change Order) and timing analysis(e.g. STA or simulation) are performed for fixing the hold violation and for verification of the Selective-MT circuit.

| Circuit | Area/Leakage | Dual-Vth | Con.-SMT | Imp.-SMT |
|---|---|---|---|---|
| A | Area | 100.00% | 164.84% | 133.18% |
| A | Leakage | 100.00% | 14.58% | 9.42% |
| B | Area | 100.00% | 142.22% | 115.65% |
| B | Leakage | 100.00% | 19.42% | 12.21% |

**Table.1: Comparison of three techniques**

Table.1 expresses our experimental results, and it shows that the standby leakage power could be reduced by 40% and the total area could be also reduced by 20%, compared with the conventional Selective-MT circuit.